\documentclass[aps,pra,longbibliography,twocolumn,floatfix,amsmath,superscriptaddress,amssymb,bibnotes]{revtex4-1}

\usepackage[english]{babel}
\usepackage[utf8]{inputenc}
\usepackage{graphicx}
\usepackage{hyperref}
\usepackage{xcolor}
\usepackage{amsmath}
\usepackage{amsfonts}
\usepackage{amssymb}
\usepackage{dsfont}
\usepackage{siunitx}
\usepackage{xcolor,import}
\usepackage{transparent}
\usepackage{todonotes}
\usepackage{sidecap} 

\definecolor{refColor}{HTML}{EA00F2}
\definecolor{figColor}{HTML}{008DF2}
\definecolor{urlColor}{HTML}{00AEF2}

\hypersetup{
    unicode=false,                 
    pdftoolbar=true,               
    pdfmenubar=true,               
    pdffitwindow=false,            
    pdfstartview={FitH},           
    pdftitle={},                   
    pdfsubject={Quantum physics},  
    pdfkeywords={},                
    pdfnewwindow=true,             
    colorlinks=true,               
    linkcolor=figColor,            
    citecolor=refColor,            
    filecolor=magenta,             
    urlcolor=urlColor              
}
\newcommand{\rs}[1]{\scriptscriptstyle \mathrm{ #1 }}

\newcommand{\bra}[1]{\left\langle #1\right|}             
\newcommand{\ket}[1]{\left| #1\right\rangle}              
\newcommand{\braket}[2]{\left\langle #1\middle|#2\right\rangle}              

\newcommand{\vc}[1]{\mathbf{#1}} 

\newcommand{\figref}[1]{Fig.\,\ref{#1}}

\begin{document}

\title{Observation of three-body correlations for photons coupled to a Rydberg superatom}

\author{Nina Stiesdal}
\affiliation{Department of Physics, Chemistry and Pharmacy, Physics@SDU, University of Southern Denmark, 5320 Odense, Denmark}
\author{Jan Kumlin}
\affiliation{Institute for Theoretical Physics III and Center for Integrated Quantum Science and Technology, University of Stuttgart, 70550 Stuttgart, Germany}
  \author{Kevin Kleinbeck}
\affiliation{Institute for Theoretical Physics III and Center for Integrated Quantum Science and Technology, University of Stuttgart, 70550 Stuttgart, Germany}
\author{Philipp Lunt}
\affiliation{Department of Physics, Chemistry and Pharmacy, Physics@SDU, University of Southern Denmark, 5320 Odense, Denmark}
\author{Christoph Braun}
\affiliation{Department of Physics, Chemistry and Pharmacy, Physics@SDU, University of Southern Denmark, 5320 Odense, Denmark}
\author{Asaf Paris-Mandoki}
\affiliation{Department of Physics, Chemistry and Pharmacy, Physics@SDU, University of Southern Denmark, 5320 Odense, Denmark}
\affiliation{Instituto de F\'{i}sica, Universidad Nacional Aut\'{o}noma de M\'{e}xico, Mexico City, Mexico}
\author{Christoph Tresp}
\affiliation{Department of Physics, Chemistry and Pharmacy, Physics@SDU, University of Southern Denmark, 5320 Odense, Denmark}
\author{Hans Peter B\"uchler}
\affiliation{Institute for Theoretical Physics III and Center for Integrated Quantum Science and Technology, University of Stuttgart, 70550 Stuttgart, Germany}
  \author{Sebastian Hofferberth}
\affiliation{Department of Physics, Chemistry and Pharmacy, Physics@SDU, University of Southern Denmark, 5320 Odense, Denmark}
\email{hofferberth@sdu.dk}
\date{\today}

\pacs{}


\begin{abstract}
  We report on the experimental observation of non-trivial three-photon correlations imprinted onto initially uncorrelated photons through interaction with a single Rydberg superatom. Exploiting the Rydberg blockade mechanism, we turn a cold atomic cloud into a single effective emitter with collectively enhanced coupling to a focused photonic mode which gives rise to clear signatures in the connected part of the three-body correlation function of the out-going photons. We show that our results are in good agreement with a quantitative model for a single, strongly coupled Rydberg superatom. Furthermore, we present an idealized but exactly solvable model of a single two-level system coupled to a photonic mode, which allows for an interpretation of our experimental observations in terms of bound states and scattering states.
\end{abstract}

\maketitle

Engineering effective interactions between individual optical photons is essential for applications in classical and quantum computation, communication and metrology \cite{Kimble2008c,Lukin2014b,Rempe2015b}.
The most established approach to achieve two-photon interaction is strong coupling of light to individual quantum emitters, either in resonators \cite{Raimond2001,Dayan2014,Lukin2014,Ritter2016} and waveguides \cite{Wallraff2013b,Rauschenbeutel2014,Sandoghdar2014,Lodahl2015c,Kimble2015b,Lukin2016d}, or through tight focussing in free-space \cite{Unlu2007,Kurtsiefer2008,Sandoghdar2016}. A complementary approach combines electromagnetically induced transparency and strong interaction between atoms in Rydberg states to convert photons into interacting Rydberg polaritons in an extended medium \cite{Adams2010,Vuletic2012,Hofferberth2016d}. Here we exploit a combination of the two concepts by using the Rydberg blockade mechanism \cite{Lukin2001c} to convert an atomic ensemble containing $N$ individual atoms into a single effective two-level quantum system, a Rydberg superatom, with strongly enhanced coupling to a single photonic mode \cite{Kuzmich2012b,
Hofferberth2017}.

A central concept for quantifying the influence of effective photon-photon interaction in any of these systems is to study the intensity correlations imprinted by the interaction onto initially uncorrelated photons by determination of $n$-body correlation functions
\begin{equation}
 g^{(n)}\big(s_{1}, \ldots, s_{n}\big)=  \frac{ \Big \langle E^{\dag}(s_1) \ldots  E^{\dag}(s_n) E(s_n)\ldots E(s_1)\Big \rangle}{ \prod_{i=1}^{n}\Big \langle E^{\dag}(s_i)   E(s_i) \Big \rangle}.
 \label{eq:corrfunc}
\end{equation}
Here, the  operators $E^{\dag}(s)$ and $E(s)$ describe the creation and annihilation of photons at time $s$. The outgoing photon rate is related to these operators via $I(s) = \langle E^{\dag}(s)E(s)\rangle $. Two-photon correlations $g^{(2)}$ have been extensively studied for a variety of systems \cite{Scully1997} and the observation of anti-bunching is accepted as a the characteristic fingerprint of a single photon source \cite{Polyakov2011}, while bound states of two photons have been observed as strong bunching feature in Rydberg polariton systems \cite{Vuletic2013}. To study interaction between three photons, it is natural to turn to third-order correlations $g^{(3)}$ \cite{Hvam2009,Rempe2011,Vuletic2018}.  While any
two-body correlation will also induce a signal in the three-body correlation function, a natural approach is to subtract these trivial contributions via the cumulant expansion to  identify the pure three-body correlations. This approach leads  to the connected part of the three-body correlation function
\begin{equation}
 g^{(3)}_{c}(s_1,s_2,s_3)= 2+g^{(3)}(s_1,s_2,s_3)  - \sum_{i<j} g^{(2)}(s_i,s_j) .
 \label{eq:g3conn}
\end{equation}
Note, that $g^{(3)}_{c}$ vanishes if one photon is separated from the other two. Furthermore, for any classical Gaussian state of photons, the connected part of the three-body correlation function is zero.

\begin{figure}[t]	
\includegraphics{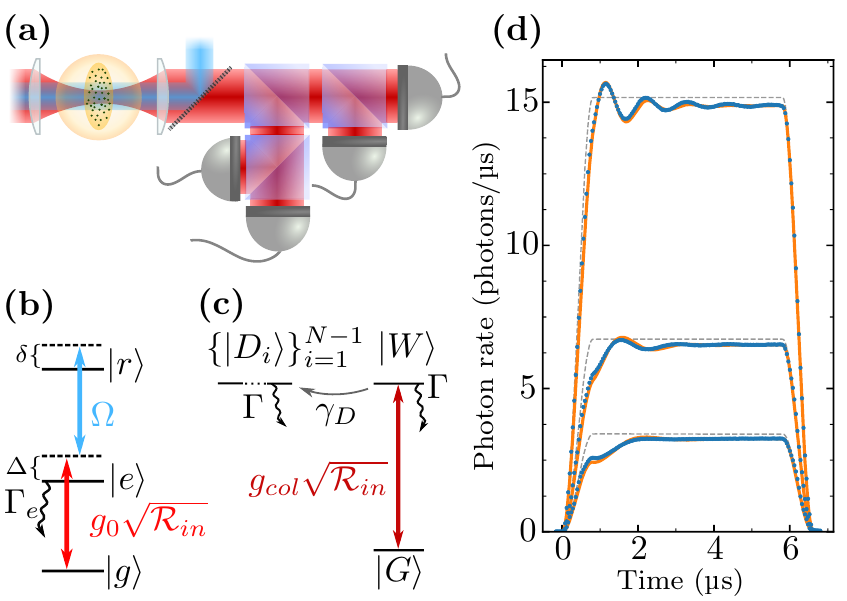}
\caption{\label{fig:Fig1} (a) Sketch of the experimental setup to couple a single Rydberg superatom to a few-photon light field. (b) Single-atom level scheme. A weak probe and a strong control field couple the ground state $\ket{g}$ via the intermediate state $\ket{e}$ to the Rydberg state $\ket{r}$. (c) Due to the Rydberg blockade the atomic ensemble behaves as a single two-level system with ground state $\ket{G}$ and collective excited state $\ket{W}$. In addition to decay, the excited state can dephase into the dark state manifold $\{\ket{D_i}\}_{i=1}^{N-1}$. (d) Time traces of the outgoing probe photon pulses (blue points) for incoming pulses (broken lines) with three different peak photon rates $\mathcal{R}_{in} = \SI{3.4}{\per\micro\per\second}$, $\mathcal{R}_{in} = \SI{6.7}{\per\micro\per\second}$, and $\mathcal{R}_{in} = \SI{15.2}{\per\micro\per\second}$. Solid orange lines show fits of our single-emitter model to the data.}
\end{figure}

Here, we report the first experimental observation of three-photon correlations by clear signatures in the connected part of the three-body correlation function of initially uncorrelated photons interacting with a single Rydberg superatom. Our setup is based on a cold atomic cloud interacting with a focused photonic mode coupling to a highly excited Rydberg state via a far detuned intermediate state. The transversal size of the photonic mode as well as the longitudinal extent of the atomic cloud are smaller than the Rydberg blockade volume resulting in a single effective emitter with collectively enhanced coupling to the probe light \cite{Hofferberth2017}. We demonstrate that such a superatom gives rise to non-trivial three-body correlations, in good agreement with a quantitative model for our system. Finally, we present an idealized but exactly solvable model of a single two-level system coupled to a photonic mode, which allows for an interpretation of our experimental observations in terms of a three-body bound state coherently superimposed with two-body bound states and scattering states. We note that recent experiments with Rydberg polaritons have also found signatures consistent with a three-body bound state of photons \cite{Vuletic2018}.

The experiment starts with trapping an ensemble of $24 \times 10^3$ laser cooled $\text{Rb}^{87}$ atoms at a temperature of $\SI{5.9}{\micro\kelvin}$ in an optical dipole trap positioned in the focus ($w_0 = \SI{6.5}{\micro\meter}$) of a weak probe beam at a wavelength of $\SI{780}{\nano\meter}$ (Fig.\,\ref{fig:Fig1}a). The extensions of the atomic medium (given by $1/e$ of the Gaussian density distribution) are $\sigma_z = \SI{6.5}{\micro\meter}$ along the probe beam and $\sigma_{x,y}=10,\SI{21}{\micro\meter}$ in the transverse directions. We couple atoms from their ground state $|g\rangle=|5S_{1/2},F=2,m_F=2\rangle$ to the Rydberg state $|r\rangle = |111S_{1/2},J=1/2,m_J=1/2\rangle$ by overlapping the probe beam with a strong counter-propagating control beam $\lambda_c=\SI{479}{\nano\meter}$ (Fig.\,\ref{fig:Fig1}b). With a single-photon detuning $\Delta = 2 \pi\times \SI{100}{\mega\hertz}$ and the two-photon detuning $\delta$ on Raman-resonance, the intermediate state $\ket{e} = \ket{5P_{3/2},F=3,m_F=3}$ can be adiabatically eliminated, reducing the dynamics of each atom to those of a resonantly driven two-level system, with the upper-state decay rate being dominated by spontaneous Raman decay via $|e\rangle$ with rate $\Gamma = \Omega^2/(2\Delta)^2 \Gamma_e$.

With a van-der-Waals coefficient of $C_6=1.88\times\SI{e5}{\giga\hertz\, \micro\metre^6}$ \cite{Hofferberth2017b} and a control field Rabi frequency $\Omega= 2 \pi \times \SI{12}{\mega\hertz}$, the radius of the Rydberg blockade volume is sufficiently large to blockade the whole intersection of probe beam and sample,so that only a single Rydberg excitation is allowed at any time. We verify this through the sub-Poissonian counting statistics of field-ionized Rydberg atoms detected on a microchannel plate. Under these circumstances, $N \approx 5\times 10^3$ atoms within the overlap volume with the probe beam couple collectively to the propagating light mode. Specifically, the $N$-body ground state $\ket{G}=\ket{g_1,\dots, g_N}$  couples to the many-body excited state $\ket{W} = \frac{1}{\sqrt{N}}\sum_{j=1}^N e^{i \vc{k}\cdot\vc{x}_j}\ket{j}$, where $\ket{j}=\ket{g_1,\dots,r_j,\dots,g_N}$ is the state with the $j$-th atom in $\ket{r}$ and all others in $\ket{g}$, $\vc{k}$ is the sum of the wavevectors of the probe and control fields and $\vc{x}_j$ denotes the position of the $j$-th atom. These two states form a single two-level superatom, which couples to the probe light with collective coupling constant $g_{\rs{col}}=\sqrt{N} g_{\rs{0}} \Omega/(2\Delta)$ (Fig.\,\ref{fig:Fig1}c). The complete basis of collective states also
contains $N-1$ dark states $\left\{\ket{D_i}\right\}_{i=1}^{N-1}$ formed by linear combinations of $\left\{\left|j\right\rangle\right\}_{j=1}^{N}$. Since all these states are orthogonal to $\ket{W}$, they do not couple to the probe light, but they still contain one Rydberg excitation blocking the medium. 
Thermal motion of the individual atoms constitutes the main dephasing from the strongly coupled state $\ket{W}$ into the collective dark states in our system. Additionally, the exchange of virtual photons between atoms provides coupling between the $\ket{W}$ state and the manifold of dark states $\left\{\ket{D_i}\right\}_{i=1}^{N-1}$ \cite{Lehmberg1970,Scully2008,Cirac2008,Molmer2013}, which can provide a fundamental limit on the minimal dephasing rate \cite{Buechler2018}.

We probe the superatom with Tukey-shaped probe pulses with a peak photon rate $\mathcal{R}_\mathrm{in}$ and collect the transmitted probe photons with a set of four single-photon counters (Fig.\,\ref{fig:Fig1}a). After each probe pulse, a field-ionization pulse removes the possibly remaining excitation in the medium, resetting the superatom to state $\ket{G}$. Fig.\,\ref{fig:Fig1}d shows time traces of the outgoing probe pulses for three different photon rates $\mathcal{R}_{in} = \SI{3.4}{\per\micro\per\second}$, $\mathcal{R}_{in} = \SI{6.7}{\per\micro\per\second}$, and $\mathcal{R}_{in} = \SI{15.2}{\per\micro\per\second}$. Because of the low total photon number in the probe pulses, the Rabi oscillation of the Rydberg population \cite{Saffman2009,Grangier2009,Kuzmich2012c} is visible in the transmitted photon stream as a periodic modulation, which results from the absorption and subsequent stimulated emission of single photons by the coherently driven two-level atom.

\begin{figure}[t]	
	\includegraphics[width=\columnwidth]{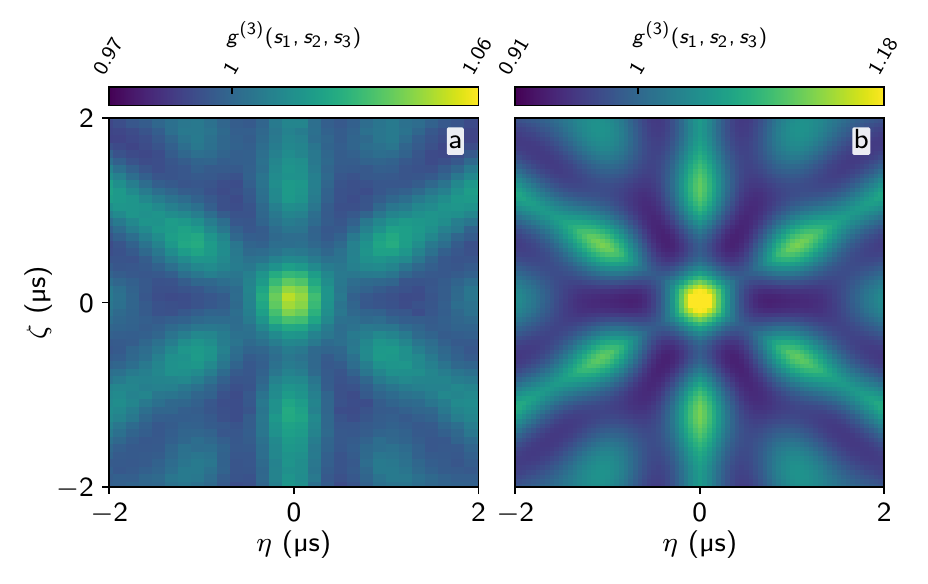}
	\caption{\label{fig:Fig2} (a) Cut through the experimental third-order correlation $g^{(3)}(s_1,s_2,s_3)$ for  input photon rate $\mathcal{R}_{in} = \SI{6.7}{\per\micro\per\second}$ along the relative Jacobi coordinates $\eta,\zeta$ averaged over the center-of-mass coordinate range $R_{\rs{range}}=\sqrt{3} \times (2.5 ...3.5)\SI{}{\micro\second}$. (b) Corresponding theoretical calculation based on the single-emitter model using the parameters extracted from the fitted curves in Fig.\,\ref{fig:Fig1}d.}
\end{figure}
On the theoretical side, the quantitative description of the experimental setup is described in detail in Ref.~\cite{Hofferberth2017}. The basic idea is to treat the superatom as
a single two-level system coupled to a quantized light field,
\begin{equation}
 H=  \int \frac{\mathrm{d}k}{2\pi} \hbar \: c k \: a_k^\dagger a_k +  \hbar \sqrt{\kappa} \left( E^\dagger (0) \sigma_{\rs{GW}}  \!+\! E(0) \sigma^\dagger_{\rs{GW}} \right) \, ,
\label{eq:Hamiltonian1}
\end{equation}
where $a_k$ and $a^\dagger_k$ are photon annihilation and creation operators, while $E(x) = \left( \sqrt{c} / (2\pi) \right) \int dk e^{ikx} a_k$
is the electric field operator measured in $\sqrt{\text{photons} / \text{time}}$, and $\sigma_{\alpha \beta} = \ket{\alpha} \bra{\beta}$.
Here, $\sqrt{\kappa}\equiv g_{\rs col}/2$ describes the collective coupling between the superatom and the light field and accounts for the
collectively enhanced spontaneous decay rate $\kappa$. Eliminating the photonic modes for a coherent input pulse,
leads to a Master equation for the reduced density matrix $\rho(t)$ of the atom alone
\begin{align}
\partial_t \rho(t) = - \frac{i}{\hbar} \left[ H_0(t), \rho(t) \right] + (\kappa + \Gamma) \mathcal{D}[\sigma_{GW}] \rho(t) \nonumber \\
+ \gamma_d \mathcal{D}[\sigma_{DW}] \rho(t) + \Gamma \mathcal{D}[\sigma_{GD}] \rho(t)
\end{align}
with the Lindblad dissipator $\mathcal{D}[\sigma]  = \sigma \rho \sigma^\dagger - (\sigma^\dagger \sigma \rho + \rho \sigma^\dagger \sigma)/2$  and the driving Hamiltonian
$H_0(t) = \hbar \sqrt{\kappa} \alpha^*(t) \sigma_{GW}  + \rm{h.c}$.
Here, the coherent field amplitude $\alpha(t)$ is related to the time-dependent mean photon rate by $\vert \alpha(t) \vert^2 = \mathcal{R}_\text{in}(t)$.  In addition to the intrinsic collectively
enhanced decay $\kappa$ into the photonic mode, we include phenomenologically the spontaneous decay of the Rydberg level with rate $\Gamma$ as well as the dephasing of the superatom state $\ket{W}$
into the manifold of dark states with rate $\gamma_d$. The outgoing electric field operator is determined by $E(t) = \alpha(t) - i \sqrt{\kappa} \sigma_{GW}(t)$.
%
Fitting these parameters by comparing the theoretical predictions with the experimental outgoing photon time traces (orange lines in Fig.\,\ref{fig:Fig1}d) yields
a single set of parameters $\kappa=\SI{0.55}{\micro s^{-1}}$,
$\Gamma=\SI{0.14}{\micro s^{-1}}$, $\gamma_d=\SI{1.49}{\per\micro\per\second}$.

Next, we calculate the third-order correlations $g^{(3)}(s_1,s_2,s_3)$ (eq.\,\ref{eq:corrfunc}) from the outgoing photon traces. A natural choice for visualization of these correlations is to transform to Jacobi coordinates $R= \left(s_1+s_2+s_3\right)/\sqrt{3}$, $\eta=\left(s_1-s_2\right)/\sqrt{2}$ and $\zeta=\sqrt{2/3}\left[\left(s_1+s_2\right)/2 - s_3\right]$. Since we are investigating the response of the superatom to a pulsed probe, the correlation function is not stationary and thus depends on the center-of-mass value $R$ as well as on the relative coordinates $\eta$ and $\zeta$. Nevertheless, this choice of coordinates lets us average the $g^{(3)}$ function over a limited range of $R$, corresponding to times $(s_1,s_2,s_3)$ within the flat-top part of the Tukey pulse. In Fig.\,\ref{fig:Fig2}a, we show the third-order correlation function $g^{(3)}(\eta,\zeta)$ for the input photon rate $\mathcal{R}_{in} = \SI{6.7}{\per\micro\per\second}$ averaged over the time range $R_{\rs{range}}=\sqrt{3}\times(2.5 ...3.5)\,\SI{}{\micro\second}$ (with respect to the time axis shown in Fig.\,\ref{fig:Fig1}d). While the averaging over $R$ certainly reduces the visibility of the three-body correlations, it is essential to extract a significant signal from the few-photon data for a realistic number of repetitions of the experiment.

Within the theoretical model described above, any multi-time correlation functions are conveniently calculated using the quantum regression theorem. Note, for a single superatom the quantum regression theorem is  exact as the emitted photons never interact with the system again \cite{Cirac2015}. The theoretical third-order correlations based on the parameters extracted from the time trace fits and averaged over the time range $R_{\rs{range}}$ are shown in \figref{fig:Fig2}b, reproducing to very good agreement the bunching and anti-bunching features observed in the experimental data.

\begin{figure}[t]
\begin{center}
\includegraphics[width=\columnwidth]{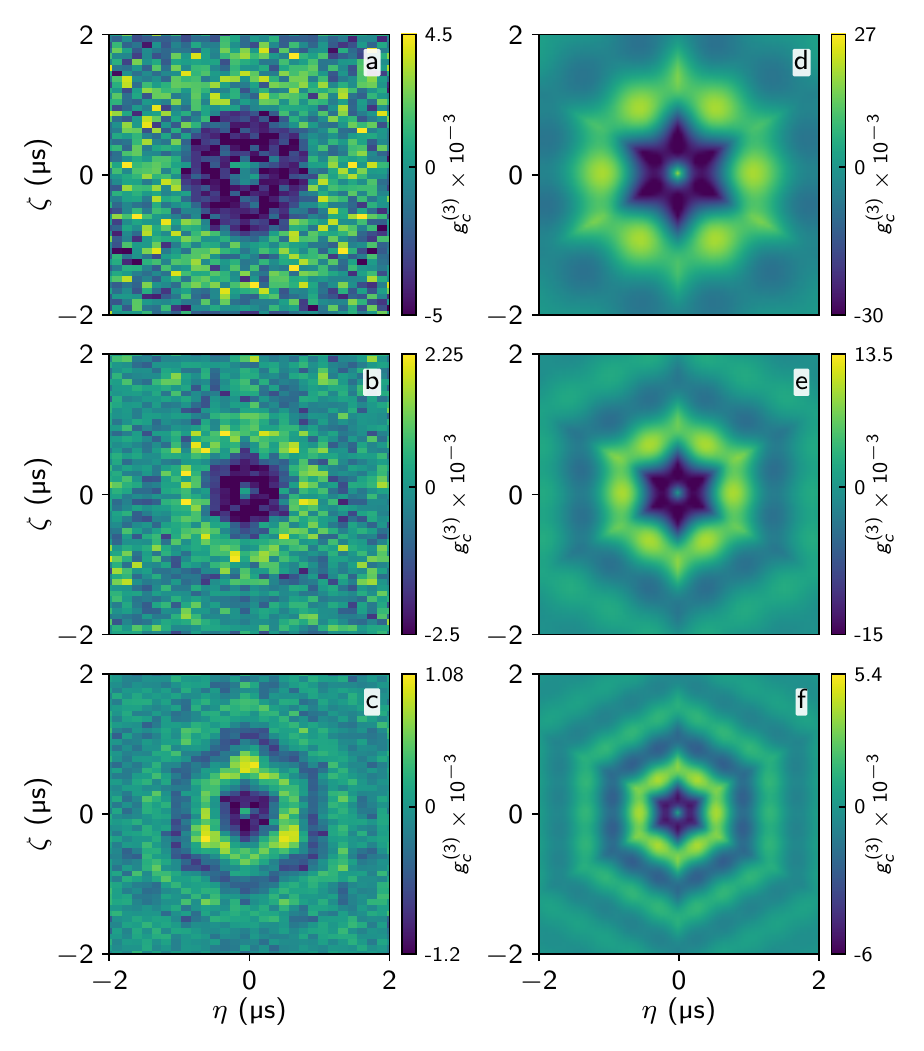}
\end{center}
\caption{Connected part of the three-photon correlation function $g^{(3)}_{c}$ in the Jacobi coordinates $\eta,\zeta$. (a), (b) and (c) show experimental results for $\mathcal{R}_{in} = \SI{3.4}{\per\micro\per\second}$, $\mathcal{R}_{in} = \SI{6.7}{\per\micro\per\second}$, and $\mathcal{R}_{in} = \SI{15.2}{\per\micro\per\second}$ respectively. (d), (e), and (f)
 show the corresponding theoretical predictions. \label{fig:Fig3}}
\end{figure}
To quantify pure three-body correlations in the outgoing photon stream we now extract the connected third-order correlation function $g^{(3)}_{c}$ by subtracting the two-body correlations for all pairwise combinations of time coordinates $s_1,s_2$ and $s_3$ from $g^{(3)}$, as defined in eq.\,\ref{eq:g3conn}. Fig.\,\ref{fig:Fig3}a-c shows cuts $g^{(3)}_{c}(\eta,\zeta)$ through the measured connected three-body correlation function for all three investigated photon input rates $\mathcal{R}_{in} = \SI{3.4}{\per\micro\per\second}$, $\mathcal{R}_{in} =\SI{6.7}{\per\micro\per\second}$, and $\mathcal{R}_{in} = \SI{15.2}{\per\micro\per\second}$ averaged over $R_{\rs{range}}$. Even for low photon numbers, we find  a clear signal of three-photon correlations in the connected part of $g^{(3)}$ with a three-photon bunching at short distances, accompanied by an anti-bunching at intermediate separations, followed by another ring of bunching. This sequence of bunching and anti-bunching features increases with increasing photon numbers. Note that $g^{(3)}$ for a translationally invariant system exhibits a six-fold symmetry in Jacobi coordinates. The reduction to a three-fold symmetry, visible in particular in Fig.\,\ref{fig:Fig3}c, is a consequence of the finite length of our probe pulse. In Fig.\,\ref{fig:Fig3}d-e we show the corresponding theoretical predictions from our quantitative model, which well-reproduce the observed structure in the experimental data.

\begin{figure}[t]
\includegraphics[width=\columnwidth]{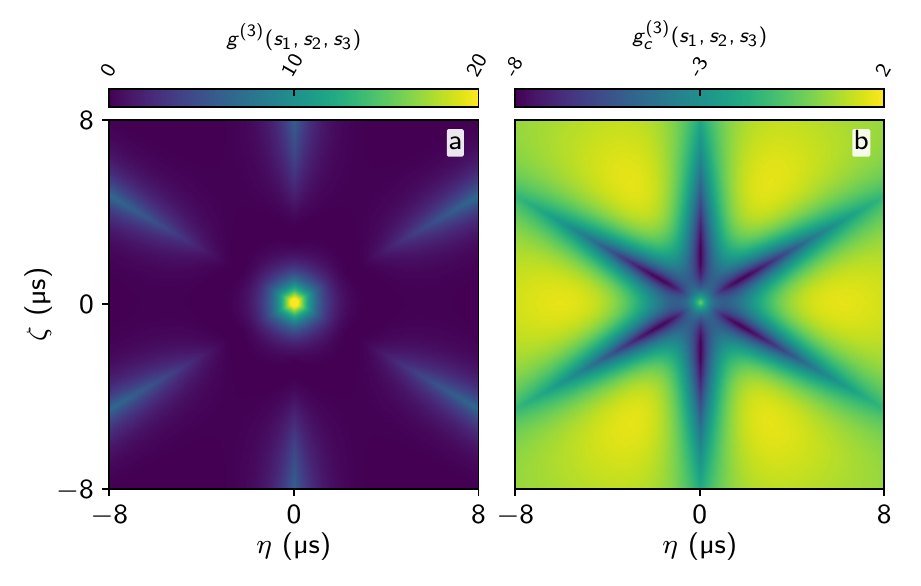}
\caption{Three-photon correlation functions derived from the wave function $\psi^{(3)}_{\rs{out}}$
 of the exactly solvable idealized model in the Jacobi coordinates $\eta,\zeta$. (a)
 The correlation function $g^{(3)}$ exhibits strong three-photon bunching at $\eta = \zeta=0$, with a subsequent anti-bunching
 and finally saturation at $g^{(2)}$ if one photon is separated from the other two.
 (b) Connected part of the correlation function $g^{(3)}_{c}$ still shows a strong non-trivial contribution.
\label{fig4}}
\end{figure}

While this agreement between theory and experiment suggests that our simple single-emitter model captures the physics of the superatom-light interaction and the effective photon-photon interaction mediated through the superatom very well, it is not straight-forward to understand the microscopic origin for the appearance of three-photon correlations from this model. In order to provide a microscopic and qualitative understanding of these correlations, we turn to the theoretical study of an idealized setup, which allows for a fully analytical solution. For this purpose, we point out that the Hamiltonian Eq.~(\ref{eq:Hamiltonian1}) is exactly solvable via the Bethe Ansatz~\cite{Yudson1985}. While such an approach ignores the additional dephasing and spontaneous emission of the excited state, it allows us to gain a microscopic understanding of the possible correlations induced by a single superatom.
A first important aspect is that the exact eigenstates for three-photons can be characterized as a three-photon bound state, a combination of a two-body bound state with an additional scattering photon, and finally pure scattering states \cite{Yudson1985}. Especially, the three-photon bound state naturally provides a non-trivial contribution to the connected part of the correlation function. A second important aspect is the relation between the outgoing wave function and the incoming photon wave function, which for the considered problem can be derived in a closed form \cite{Yudson2008}. For the present setup, we are interested in an incoming state $\psi_{\rs{in}}= \prod_{i}\psi(s_{i})$ for $n$-photons in the single photon mode $\psi$ with width $\tau$. Then, the outgoing $n$-photon wave function after interacting with a single superatom at times $s_{1} \geq  \ldots \geq s_{n}$ reduces to (see supplement material)
\begin{eqnarray}
\psi^{(n)}_{\rs{out}}(s_1,\ldots,s_n) &  = & \partial_{\alpha_1} \ldots \partial_{\alpha_n}
 e^{\frac{\kappa}{2} \sum_{i} \left(s_{i}-2\alpha_{i}\right)}   \\
&& \hspace{-50pt}
\times  \phi(s_1-\alpha_1)
 \prod_{i=2}^{n}\Big[\phi(s_{i}\!-\!\alpha_{i}) -  \phi(s_{i\!-\!1}\!+\!\alpha_{i\!-\!1})  \Big]\bigg|_{\alpha_{i}=0} \nonumber
\end{eqnarray}
with $\phi(s)=\int_{s}^{\infty}dt \psi(t) e^{-\kappa t/2} $; the wave function for all values of $\{s_{i}\}$ is obtained by requiring the bosonic symmetry of the wave function.
For three-photons in a wide incoming mode $\tau \kappa \gg 1$, the wave function in the center of the pulse reduces to ($s_{1} \geq  s_{2}\geq s_{3}$)
\begin{equation}
 \psi^{(3)}_{\rs{out}} = 1+ 12 e^{- \kappa \frac{s_{1}-s_{3}}{2}} - 4 \left(e^{- \kappa\frac{s_{1}-s_{2}}{2}}+e^{- \kappa \frac{s_{2}-s_{3}}{2}}\right),
\end{equation}
from which we can analytically derive the three-body correlation functions shown in Fig.\,\ref{fig4}. We find a very strong three-photon bunching around $\eta=\zeta=0$. The contribution of the three-body bound state to this signal can be observed by the decomposition of the wave function $\psi^{(3)}_{\rs{out}}=4 e^{- \kappa (s_{1}-s_{3})} + \psi_{\rs{sc}}$. The first term describes the three-body bound state, while $\psi_{\rs{sc}}$ accounts for the remaining contributions of scattering states and two-photon bound states with $\psi_{\rs{sc}} \!=\! 1$ at $s_1\!=\!s_2\!=\!s_3$. Therefore, the three-body bound state provides the dominant contribution to the three-photon bunching signal, but contributions of the remaining states are still significant. Especially, the exponential decrease of the bunching signal at $s_{2}\!=\!s_{3}$ exhibits a decay $\kappa$ instead of the faster decay $2 \kappa$ expected from the three-body bound state wave function.

Comparing this result with the experimentally observed correlation functions, we find that the idealized setup exhibits the characteristic features of three-photon bunching at short distances and anti-bunching at intermediate distances observed in the experiment for low photon numbers.  We expect therefore, that the microscopic origin of the three-body correlations in the experiment are well captured by an understanding of the idealized model as the combination of a three-photon bound state, scattering states and two-photon bound states. Note, that the width of these signals are increased due to the reduced losses in the idealized setup. Furthermore, the appearance of oscillations for higher number of photons can be understood as the single emitter undergoing Rabi oscillations, which gives rise to a characteristic beating for increasing photon number.

In conclusion, the experimental observation of three-photon correlations imprinted by a single Rydberg superatom on an initially uncorrelated photonic state is well accounted for by a single-emitter model, where the dephasing of the collectively excited state as well as the spontaneous decay are included. On a microscopic level, the appearance of the three-photon correlations are well understood in an idealized setup, which suggests that the observed three-photon bunching signal here cannot be purely attributed to a three-photon bound state in contrast to recent observations on Rydberg polaritions \cite{Vuletic2018}.
Observing such three-photon correlations imprinted by a single two-level systems sheds light on the fundamental processes of absorption and emission at the quantum level and highlights the potential for experimentally realizing photonic strongly correlated many-body systems in quantum nonlinear optical systems. 
Besides further improving the emitter-light coupling and investigating intrinsic superatom dephasing mechanisms, we envision the scaling of our system to more complex arrangements of multiple superatoms for implementing quantum optical networks \cite{Kimble2008c,Rempe2015b}.

\begin{acknowledgments}
We thank Florian Christaller and Simon W. Ball for contributing to the construction of the experimental setup. This work is supported by the European Union under the ERC consolidator grants SIRPOL (grant N. 681208) and RYD-QNLO (grant N. 771417), and the Deutsche Forschungsgemeinschaft (DFG) under SPP 1929 GiRyd project N. HO 4787/3-1. A.P-M. acknowledges support from  UNAM-PAPIIT IA101718 RA101718
\end{acknowledgments}

\onecolumngrid
\appendix
\section{Diagonalization of the model Hamiltonian}
The Hamiltonian Eq.~3 in main text, which describes the interaction of the photons with an
effective two level system, is an example of a quantum integrable system and is solvable by the
Bethe ansatz. The $N$-particle wave functions with energy $E_\lambda = \sum_{i=1}^N \lambda_i$
reads~\cite{Yudson1985}
\begin{align}
	\ket{\lambda} =& C(\lambda) \int\!\mathrm{d}^Ns\ \prod_{i<j}
		\left( 1 + \frac{i \kappa\ \mathrm{sgn}(s_i-s_j)}{\lambda_i-\lambda_j} \right)
		\prod_{i=1}^N f(s_i,\lambda_i) e^{i\lambda_i s_i} r^\dagger(s_i,\lambda_i)\ket{0}
	\label{eq:BetheStates}
\end{align}
where
\begin{align*}
	f(s,\lambda) &= \frac{\lambda - i\kappa/2\ \mathrm{sgn} s}{\lambda + i\kappa/2}, \\
	r^\dagger(s,\lambda) &= E^\dagger(s) + \frac{\sqrt{\kappa}}{\lambda} \delta(s) a^\dagger
\end{align*}
and $C(\lambda)$ is a normalization constant. The parameters
$\lambda = (\lambda_1,\dots,\lambda_n)$ are called rapidities. For real $\lambda$ they correspond
to momenta of incoming plane waves. However, taking only real values for $\lambda$ into account
would not suffice to build a orthonormal basis for the Hamiltonian Eq.~3 in the main text.
Hence, we need to allow complex values for $\lambda_i$. The only possibility for which
$\ket{\lambda}$ remains bounded is if $\mathrm{Re}(\lambda_i) = \mathrm{Re}(\lambda_{i+1})$ and
$\mathrm{Im}(\lambda_i) + i\kappa = \mathrm{Im}(\lambda_{i+1})$ for a subset
$\{\lambda_i,\dots,\lambda_{i+m}\}$ of rapidities. Since the energy expectation value of
$\ket{\lambda}$ must remain a real number this subset must be symmetric about the real axis.

These sets of rapidities are called strings and they are best interpreted as bound states.
For example, a two photon bound state contains rapidities
$\lambda_i = \lambda - i\kappa/2$, $\lambda_{i+1} = \lambda + i\kappa/2$ and a three photon
bound state contains $\lambda_{i} = \lambda - i\kappa$, $\lambda_{i+1} = \lambda$ and
$\lambda_{i+2} = \lambda + i\kappa$. The completeness relation expressed by the Bethe states
now reads
\begin{equation}
	\mathrm{Id} =
		\sum_{\mathrm{strings}} \int_\Lambda\!\mathrm{d}^N\lambda\,\ket{\lambda}\bra{\lambda}
	\label{eq:BetheIdentity}
\end{equation}
with $\Lambda$ the respective manifold for each string configuration.

To study scattering processes one best decomposes an initial state $\ket{\Psi_0}$ into Bethe
states and studies the time evolution in this basis. Yet, the actual calculation of the overlap
integral of $\ket{\Psi_0}$ with a given string configuration and the following summation over
all possible configurations is quite cumbersome in practice. Thus, it comes as a surprise that
the string summation can be avoided completely. Instead of integration along all possible
strings one introduces a complex contour $\Gamma$, such that
\begin{equation}
	\ket{\Psi} = \int_\Gamma\mathrm{d}^N\lambda\,\ket{\lambda}\braket{\lambda^{(A)}}{\Psi}.
	\label{eq:YudsonIdentity}
\end{equation}
For this, one needs to analytically continue the states $\ket{\lambda}$ onto
$\lambda \in \mathbb{C}^N$ and introduces the auxiliary state
\begin{equation}
	\ket{\lambda^{(A)}} = \frac{\sqrt{N!}}{(2\pi)^{N/2}} \int\mathrm{d}^Ns\,
		\theta(s_1 \ge \dots \ge s_N) \prod_{i=1}^N f(s_i, \lambda_i) e^{i \lambda_i s_i}
		r^\dagger(s_i,\lambda_i) \ket{0}.
\end{equation}
A detailed analysis and the conditions on $\Gamma$ are found in~\cite{Yudson1985}.

Using $\ket{\Psi} = \ket{s_1,\dots,s_N}$, with $0 > s_1 > \dots > s_N$, the
representation~\eqref{eq:YudsonIdentity} yields the Green's function
\begin{align}
	G(t,s) =&\ \theta(t_1 \ge s_1 \ge \dots \ge t_N \ge s_N)
	\sum_{\sigma \in S'_N} \prod_{i=1}^N
	\big[\delta(t_i - s_{\sigma_i}) - \kappa\theta(t_i - s_{\sigma_i})\big]
	e^{-\kappa(t_i - s_i)/2}
	\label{eq:GreensFunction}
\end{align}
for the asymptotic scattering outcome. The summation runs over a restricted part of the
symmetric group $S_N$ which flows from the Heaviside function in~\eqref{eq:GreensFunction}. Note
that this allows also to replace $S'_N$ by $S_N$.

\section{Generating Functional for outgoing states} The Green's function~\eqref{eq:GreensFunction}
allows in principle to calculate the photon wave function after the scattering process for any given initial photon wave function. Practically, the large number of permutations hampers such
attempts. However, we show that the Green's function owns a more compact formulation consisting
of $N$ auxiliary parameters $\alpha_i$, which in turn enables us to derive a generating
functional for the scattered wave function in the case of an initial product wave function.
For this, one uses the identity
\begin{align}
	\notag
	&\theta(t_1 \ge s_1 \ge \dots \ge t_N \ge s_N)
		\sum_{\sigma \in S'} \prod_{i=1}^N \left\{ \delta(t_i - s_{\sigma_i})
		- \kappa\theta(t_i - s_{\sigma_i}) \right\} \\
	&= \lim_{t_{N+1} \to -\infty}\prod_{i=1}^N \partial_{\alpha_i} e^{-\kappa\alpha_i}
		\theta(t_i + \alpha_i - s_i)
		\theta(s_i + \alpha_i - t_{i+1}) \bigg|_{\alpha_i = 0},
	\label{eq:OurGreen}
\end{align}
which holds almost everywhere. The proof uses induction in $N$ and is straight-forward, yet
quite cumbersome. Therefore, we will only explicitly show the case $N = 2$ here.

We calculate the right hand side of~\eqref{eq:OurGreen}, but ignore the $t_3 \to -\infty$
limit for the moment
\begin{align*}
	&\prod_{i=1}^2\partial_{\alpha_i} e^{-\kappa\alpha_i}
		\theta(t_i + \alpha_i - s_i)\theta(s_i + \alpha_i - t_{i+1}) \bigg|_{\alpha_i = 0} \\
	&= \theta(t_3 \le s_2 \le t_2 \le s_1 \le t_1)
		\left[ -\kappa + \delta(t_1 - s_1) + \delta(s_1 - t_{2}) \right]
		\left[ -\kappa + \delta(t_2 - s_2) + \delta(s_2 - t_{3}) \right].
\end{align*}
Let us now consider each term separately. First of all, we use $\delta(s_2 - t_3) = 0$ in the
limit $t_3 \to -\infty$. We start with the constructive example of the terms in
$\mathcal{O}(\kappa^2)$. Here we have
\begin{align*}
	&\theta(t_3 \le s_2 \le t_2 \le s_1 \le t_1) \kappa^2 \\
	&= \theta(t_3 \le s_2 \le t_2 \le s_1 \le t_1) \kappa^2
		\left[ \theta(t_1 - s_2)\theta(t_2 - s_2) + \theta(t_1 - s_2)\theta(t_2 - s_1) \right],
\end{align*}
since $\theta(t_1 - s_1)\theta(t_2 - s_2)$ equals one if
$\theta(t_3 \le s_2 \le t_2 \le s_1 \le t_1) = 1$, while $\theta(t_1 - s_2)\theta(t_2 - s_1)$
vanishes under the same condition, but on the zero measure set $\{s_1 = t_2\}$.
We repeat this strategy for every other power of $\kappa$, i.e., we add Heaviside- and
Delta-functions, which are either trivial --- due to the
$\theta(t_3 \le s_2 \le t_2 \le s_1 \le t_1)$ term --- or vanish everywhere, except on a set with
measure zero.

The terms in power $\mathcal{O}(\kappa)$ are
\begin{align*}
	&\theta(t_3 \le s_2 \le t_2 \le s_1 \le t_1)
	\left[ \delta(t_1 - s_1) + \delta(t_2 - s_1) + \delta(t_2 - s_2) \right] \\
	&=\theta(t_3 \le s_2 \le t_2 \le s_1 \le t_1)
		\big[ \delta(t_1 - s_1)\theta(t_2 - s_2) + \delta(t_2 - s_1)\theta(t_1 - s_2)
		+ \theta(t_1 - s_1)\delta(t_2 - s_2) + \theta(t_2 - s_1)\delta(t_1 - s_2) \big].
\end{align*}
The first three inserted Heaviside-functions are again trivial, while the last term vanishes except
on the set $\{(s_1,s_2) | s_1 = t_2\}$. The remaining terms in $\mathcal{O}(1)$ are
\begin{align*}
	&\theta(t_3 \le s_2 \le t_2 \le s_1 \le t_1)
		\left[ \delta(t_1 - s_1)\delta(t_2 - s_2) + \delta(t_2 - s_1)\delta(t_2 - s_2) \right] \\
	&=\theta(t_3 \le s_2 \le t_2 \le s_1 \le t_1)
		\left[ \delta(t_1 - s_1)\delta(t_2 - s_2) + \delta(t_2 - s_1)\delta(t_1 - s_2) \right].
\end{align*}
We used that the set with $s_1 = s_2$, i.e., where both photons at the same position, has again
vanishing measure. The other term $\delta(t_2 - s_1)\delta(t_1 - s_2)$ is zero, since $s_2$ will never
take the value $t_1$.

In total, we now have
\begin{align*}
	&\lim_{t_3\to -\infty} \left[ -\kappa + \delta(t_1 - s_1) + \delta(s_1 - t_{2}) \right]
		\left[ -\kappa + \delta(t_2 - s_2) + \delta(s_2 - t_{3}) \right] \\
	&= \theta(s_2 \le t_2 \le s_1 \le t_1) \sum_{\sigma\in S_2'}\prod_{i=1}^2
		\{\delta(t_i - s_{\sigma_i}) - \kappa\theta(t_i - s_{\sigma_i})\},
\end{align*}
Bringing us directly to the left hand side of \eqref{eq:OurGreen}, which ends
the induction basis.

The induction step uses the restriction on $S'_N$: for every $\sigma \in S'_N$ we have
$\sigma_i \ge i-1$. Thus we can split $S'_{N+1}$ into two sets, namely $(N+1,N+1)\otimes S'_N$
and $(N+1,N)\otimes S'_N$. With that we can split the summation in~\eqref{eq:OurGreen} into
these two corresponding parts and use the induction assumption.

This brings us to one of our most important results from the theory side. For an initial
product wave function $\Psi_\mathrm{in} = \prod_i \Psi(s_i)$ we find that the outgoing
wave function
\begin{align*}
	\Psi_\mathrm{out}(s) &= \int\!\mathrm{d}^Nt\, G(s,t) \Psi_\mathrm{in}(t)
\end{align*}
primarily consists of terms of the form
\begin{align*}
	&\int_a^b\!\mathrm{d}t\, e^{-\kappa t/2} \Psi(t)
	= \int_a^\infty\!\mathrm{d}t\, e^{-\kappa t/2} \Psi(t)
		-\int_b^\infty\!\mathrm{d}t\, e^{-\kappa t/2} \Psi(t).
\end{align*}
Consequently, the definition $\Phi(s) = \int_s^\infty\!\mathrm{d}t\, e^{-\kappa t/2} \Psi(t)$
yields the representation Eq.~5 from the main text for the outgoing wave function.

\twocolumngrid
%

\end{document}